\documentclass[aps,prb,floatfix,superscriptaddress,twocolumn,footinbib]{revtex4-1}
\usepackage{amssymb}
\usepackage{amsmath}
\usepackage{amsfonts}
\usepackage{appendix}
\usepackage{graphicx}
\usepackage{epsfig}
\usepackage{epstopdf}
\usepackage{balance}
\usepackage[dvipsnames]{xcolor}
\usepackage{calc}
\usepackage{natbib}
\usepackage[colorlinks,
linkcolor=blue,
anchorcolor=blue,
citecolor=blue,
urlcolor=blue]{hyperref}
\usepackage{lipsum}
\usepackage[normalem]{ulem}

\usepackage{graphicx}
\usepackage{dcolumn}
\usepackage{bm}

\begin{document}

\preprint{APS/123-QED}

\title{Valley-selective  Floquet Chern flat bands in twisted multilayer graphene}
\affiliation{International Center for Quantum Materials, School of Physics, Peking University, Beijing 100871, China}
\affiliation{Beijing Academy of Quantum Information Sciences, Beijing 100193, China}

\affiliation{Center for Advanced Quantum Studies, Department of Physics, Beijing Normal University, Beijing 100875, China}
\affiliation{School of Physics and Wuhan National High Magnetic Field Center,
	Huazhong University of Science and Technology, Wuhan 430074,  China}
\affiliation{CAS Center for Excellence in Topological Quantum Computation, University of Chinese Academy of Sciences, Beijing 100190, China}

\author{Ming Lu}
\affiliation{International Center for Quantum Materials, School of Physics, Peking University, Beijing 100871, China}
\affiliation{Beijing Academy of Quantum Information Sciences, Beijing 100193, China}
\author{Jiang Zeng}
\affiliation{International Center for Quantum Materials, School of Physics, Peking University, Beijing 100871, China}
\author{Haiwen Liu}
\affiliation{Center for Advanced Quantum Studies, Department of Physics, Beijing Normal University, Beijing 100875, China}
\author{Jin-Hua Gao}
\email{jinhua@hust.edu.cn}
\affiliation{School of Physics and Wuhan National High Magnetic Field Center,
	Huazhong University of Science and Technology, Wuhan 430074,  China}
\author{X. C. Xie}
\email{xcxie@pku.edu.cn}
\affiliation{International Center for Quantum Materials, School of Physics, Peking University, Beijing 100871, China}
\affiliation{Beijing Academy of Quantum Information Sciences, Beijing 100193, China}
\affiliation{CAS Center for Excellence in Topological Quantum Computation, University of Chinese Academy of Sciences, Beijing 100190, China}

\date{\today}

\begin{abstract}
We show that Floquet engineering with circularly polarized light (CPL) can selectively split the valley degeneracy of a twisted multilayer graphene (TMG), and thus generate a controlled valley-polarized Floquet Chern flat band with tunable large Chern number.  It offers a feasible optical way to manipulate the valley degree of freedom in moir\'{e} flat bands, and hence opens new opportunities to study the valleytronics of mori\'{e} flat band systems. We thus expect that many of the valley-related properties of TMG, \textit{e.g.} orbital ferromagnetism,  can be switched by CPL with proper doping. We reveal a Chern number hierarchy rule for the Floquet flat bands in a generic (M+N)-layer TMG. We also illustrate that the CPL effects on TMG strongly rely on the stacking chirality, which is an unique feature of TMG. All these phenomena could be tested in the twisted double bilayer graphene systems, which is the simplest example of TMG and has already been realized in experiments.

\end{abstract}

\maketitle


\section{introduction}
The topological flat bands in twisted multilayer graphene (TMG) have drawn great reasearch interest very recently, because that a Chern flat band is believed to be a promising platform to realize fractional Chern insulator and may harbour novel correlation states \cite{cao2020,liu2020tunable,cheng2020, Burg2019,Zhang2019a,Koshino2019, Chebrolu2019,Liu2019, joon2019, Lee2019,Wu2019b, Chen2019b,Chen2019a,Chittari2019,Zuo2018,Morell2013,Ma2021,Bistritzer2011,Li2019a,Carr2020,Ma2020,liujianpeng2020,shiyanmeng2020,Park2020,Polshyn2020,Chen2020}. The most generic situation is a (M+N)-layer TMG, where two ABC-stacked multilayer graphene (MG) are stacked on top of each other with a small twist angle \cite{Liu2019}. Twisted double bilayer graphene (TDBG) is the simplest example, \textit{i.e.}, the case of $M=2$, $N=2$, and has already been realized in experiments \cite{cao2020,liu2020tunable,cheng2020,Burg2019}. In TMG, a pair of flat bands are formed around the magic angle, which locate at two inequivalent valleys in momentum space.  The two flat bands can be isolated by a vertical electric field and have nonzero valley Chern numbers \cite{Zhang2019a,Koshino2019,Chebrolu2019,joon2019,Lee2019,Liu2019}. The topological flat band  is an unique feature of TMG. Another special characteristic of TMG is that it has a new degree of freedom,  \textit{i.e.}~the stacking chirality. For example, in TDBG, the AB-AB and AB-BA configurations have distinct stacking chirality arrangement, but very similar band structures \cite{Zhang2019a,Koshino2019, Chebrolu2019,Liu2019}. Neither the isolated topological flat band nor the stacking chirality is present in the celebrated twisted bilayer graphene (TBG).

However, the total Chern number of the topological flat bands in TMG is always zero due to time reversal symmetry\cite{Koshino2019, Chebrolu2019, Liu2019}. A possible improvement is via Floquet engineering using circularly polarized light (CPL), which can break time reversal symmetry and  effectively produce non-equilibrium topological phases \cite{Oka2009, Kitagawa2011, Gu2011, Wang2013, Usaj2014, Eckardt2015, Mikami2016, Chan2016, Zhang2016a, Oka2019}.  Intriguingly, several recent works studied Floquet engineering on twisted bilayer graphene systems. A tight binding model study of TBG well above the magic angle shows that the band topology induced by CPL is very similar to the monolayer graphene limit \cite{Topp2019}. Continuum model studies near the magic angle have found that the flat band width can become even smaller by CPL irradiation~\cite{Li2020a, Katz2020}, and the interlayer coupling can be tuned by a transverse magnetic mode wave at the exit of a waveguide~\cite{Vogl2020a}. Beyond the off-resonant high frequency limit, effective Hamiltonian for TBG in the low and intermediate frequency regime has been derived, where different symmetry breaking phases have been found ~\cite{Vogl2020}. 
 It is natural to expect that CPL may induce more complicated and interesting phenomena in TMG, due to its complex structure and novel topological properties. More interestingly,
in valleytronics, CPL is a  rather effective way  to manipulate the valley degree of freedom in honeycomb lattices, such as bilayer graphene,  $\textrm{MoS}_2$, \textit{etc} \cite{xiaodi2007,Rycerz2007, apl2009,bilayer2011,cao2012,xiaodi2012,jin2014,  Marino2015, xuxiaodong2016,babak2016,sie2017}. Thus, whether the CPL is able to give rise to some valley-related phenomena in TMG is also a worthwhile question, since the studies about how to manipulate the valley degree of freedom of moir\'{e} flat bands is still rare. 

In this work, we study the CPL irradiation  induced   Floquet-Bloch band sturctures of the TMG systems. We show that, near the charge neutrality point, TMG  always has two isolated Floquet Chern flat bands in the presence of CPL irradiation. Most importantly, we find that with the help of CPL irradiation, we can selectively split the valley degeneracy of TMG, and thus generate  an optical controlled valley-polarized Floquet Chern flat band with tunable large Chern number.  We reveal the Chern-number hierarchy rules of the Floquet flat bands in the (M+N)-layer TMG, which strongly relies on the stacking chirality of TMG. An intuitive explanation about why the TMG with different stacking chiralities has distinct responses to CPL is also given. Since CPL is able to control the valley polarization of the Floquet flat bands in TMG, we expect that many of the valley-related properties of TMG~\cite{liujianpeng2020}, like orbital ferromagnetism, quantum anomalous Hall effect, magneto-optical and nonlinear optical propertieis,  can be switched by CPL.  Our work thus opens up new opportunities to study the valleytronics of moir\'{e} flat bands.

\begin{figure}[tbp]
\includegraphics[scale=0.27]{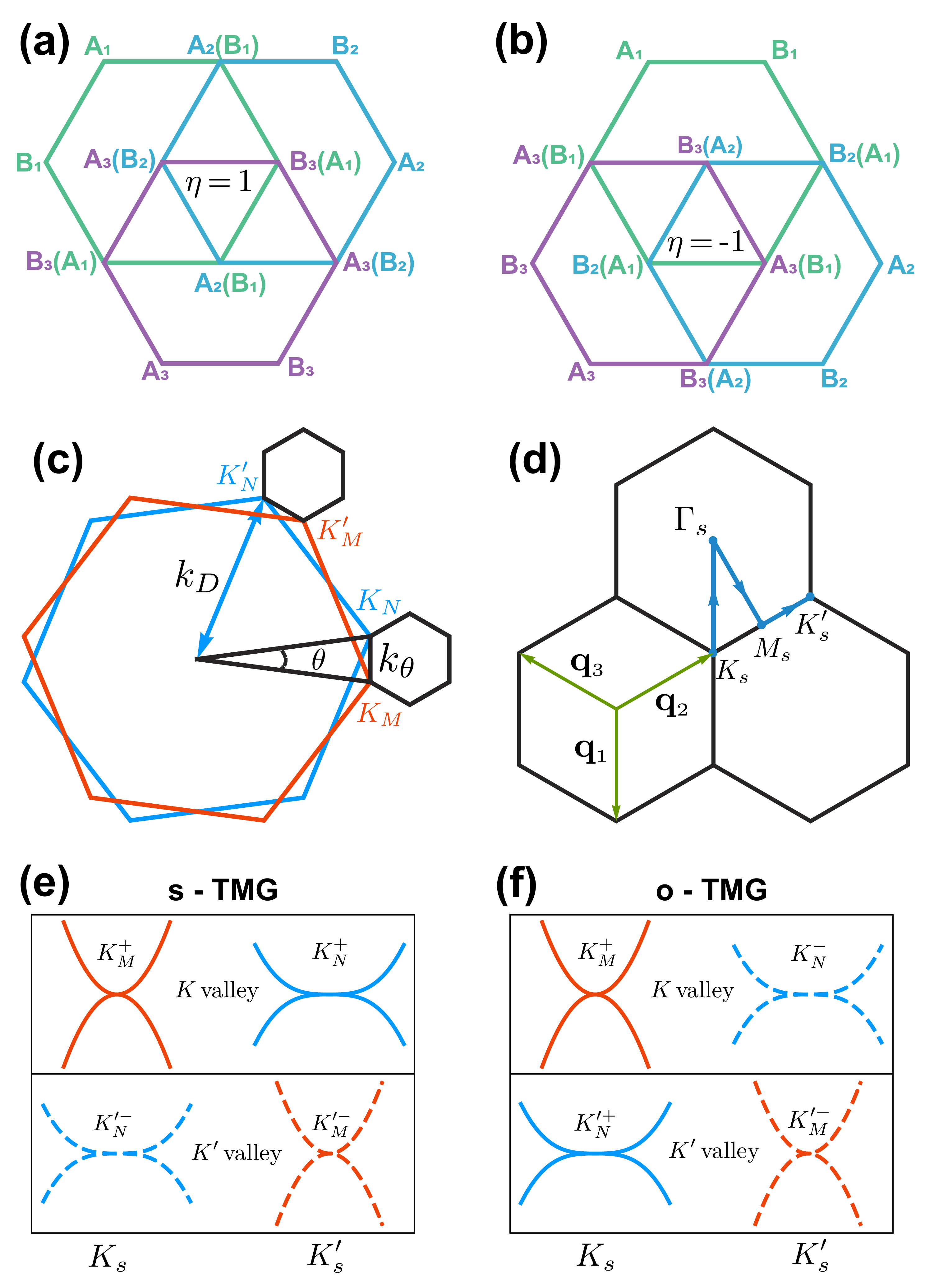}
\caption{\label{fig:fig1}(a) and (b) are respectively the schematics the chirally stacked MGs with $\eta=\pm 1$ stacking chirality. Here, $A_2$ means A site in layer 2, and so on. (c) Brillouin-zone (BZ) of the top (M-layer, red) and bottom (N-layer, blue) MGs. The black hexagon represent the  mori\'e Brillouin-zone (mBZ) of TMG.  (d) A larger version of mBZ. (e) and (f) illustrate the valley chirality of the MGs for each moir\'{e} valley. The case when two MGs have the same stacking chirality, like AB-AB TDBG, is given in (e); while (f) is for the opposite case, \textit{e.g.}~AB-BA TDBG.}
\end{figure}

\section{Model Hamiltonian}
We consider a (M+N)-layer TMG in the presence of CPL irradiation. In Figs.~\ref{fig:fig1}(a) and \ref{fig:fig1}(b), we first illustrate the two distinct stacking chiralities of the ABC-stacked MG \cite{min2008,zhangfan2010,jung2011}, where the ABC (CBA) configuration is denoted as $\eta=1$ ($\eta=-1$). Here, $\eta$ is the index of stacking chirality. Thus, TMG can be divided into two categories: s-TMG where two MGs have same stacking chirality (like AB-AB TDBG) and o-TMG with opposite stacking chirality (like AB-BA TDBG).


We consider the TMG with small twist angle $\theta$ near the first magic angle \cite{Santos2007,Santos2012,Bistritzer2011,Liu2019,Koshino2019,Chebrolu2019}. The corresponding moir\'{e} Brillouin zone (mBZ) (black lines) is given in Fig.~\ref{fig:fig1} (c), where red (blue) line is the BZ of the top M-layer (bottom N-layer) MG. We see that the two MGs give a pair of $K$ valley ($K_M$ and $K_N$) and a pair of $K'$ valley ($K'_M$ and $K'_N$).   Meanwhile,  the moir\'{e} interlayer hybridization only mix the adjacent two valleys near either $K$ or $K'$, while the interaction between distant valleys are tiny.  Thus, valley is also a good quantum number of TMG, which are denoted as $K$ ($\tau=+1$) and $K'$ ($\tau=-1$) moir\'{e} valley. Here, $\tau$ is the moir\'{e} valley index.

A notable feature is the chirality of MG valleys, which will significantly influence the CPL effects as shown later. We know that the two inequivalent valleys in MG have opposite chirality, which also rely on its stacking chirality. Without loss of generality, we fix the stacking chirality of the top M-layer MG to be $\eta=+1$ and define the chirality of its $K$ valley as ``+", which is denoted as $K_M^+$. Then, for the s-TMG, the moir\'{e} valley $K$ ($K'$) is composed of $K^+_M$  and $K^+_N$ ($K'^-_M$ and $K'^-_N$) , as shown in Fig.~\ref{fig:fig1} (e).  In contrast,  moir\'{e} valley $K$ ($K'$) of the o-TMG corresponds to   $K^+_M$ and $K^-_N$ ($K'^-_M$ and $K'^+_N$), see Fig.~\ref{fig:fig1} (f).

We assume a normal incident CPL, which is described by a vector potential $\bm{A}(t)=A_0(\cos\Omega t, -\xi\sin\Omega t)$. Here, $\xi=1$ ($\xi=-1$) represents  a left (right) CPL, and  $A_0$, $\Omega$  are the amplitude  and frequency, respectively. The time dependent Hamiltonian of the irradiated (M+N)-layer TMG in $K$ valley is
\begin{equation}\label{TimeH}
    H_{K,M+N}^{\eta,\eta'}(t)=\begin{pmatrix}
    H_{K,M}^{\eta}(t)& \mathbb{T}\\
    \mathbb{T}^\dagger & H_{K,N}^{\eta'}(t)
    \end{pmatrix} + U,
\end{equation}
where  $H_{K,M}^{\eta}(t)$ and $H_{K,N}^{\eta'}(t)$ describe the top and bottom MGs, respectively.  $\mathbb{T}$ represents the moir\'{e} interlayer hopping, and $U$ models the gate induced  potential difference between layers.
Specifically,
\begin{equation}
    H_{K,M}^{\eta}=\begin{pmatrix}
    h_0(t)& h_{\eta} & 0 & \cdots\\
    h_{\eta}^\dagger & h_0(t) & h_{\eta} & \cdots \\
    0 & h_{\eta}^\dagger & h_0(t) & \cdots \\
    \vdots & \vdots & \vdots & \ddots
    \end{pmatrix},
\end{equation}
where $h_0(t)=v_F\sigma\cdot\left[-i\hbar \nabla+e A(t)\right]$ is the Hamiltonian of monolayer graphene in $K$ valley. The stacking chirality is reflected in the interlayer hopping matrix $h_{\eta}$, where
$
    h_{\eta=+}=\begin{pmatrix}
    0 & 0\\
    t_\bot & 0
    \end{pmatrix}
$ and $h_{\eta=-}=h_{+}^{\dagger}$ with $t_\bot$ being the nearest neighbor interlayer hopping.  $\mathbb{T}=E_{M\times N}\otimes T(r)$,  $E_{M\times N}$ is a $M\times N$ matrix with only one nonzero entry $E_{M\times N}(M, 1)=1$.
$T(r)=\sum_{n=1}^3 T_n e^{-i\bm{q}_n\cdot \bm{r}}$ is the twist tunneling matrix of TMG, where
 $\bm{q}_{n+1}=k_\theta(\sin\, n\phi, -\cos\, n\phi)$, $\phi=2\pi/3$,  $k_\theta=2k_D\sin(\theta/2)$, see in Figs.~\ref{fig:fig1} (c) and (d). $T_{n+1}=w_{\mathrm{AA}}\sigma_0+w_{\mathrm{AB}}(\sigma_x \cos n\phi+\sigma_y\sin n\phi)$, where $w_{\mathrm{AA}}$ ($w_{\mathrm{AB}}$) represents the tunneling amplitude of the intra- (inter-) sublattice.   $U=\mathrm{diag}(M-\frac{1}{2},\cdots, \frac{1}{2}, -\frac{1}{2}, \cdots, -N+\frac{1}{2})\otimes \Delta_E\sigma_0$,  where $\Delta_E$  is the gate induced potential difference between adjacent layers.

\begin{figure}[tbp]
\includegraphics[scale=0.52]{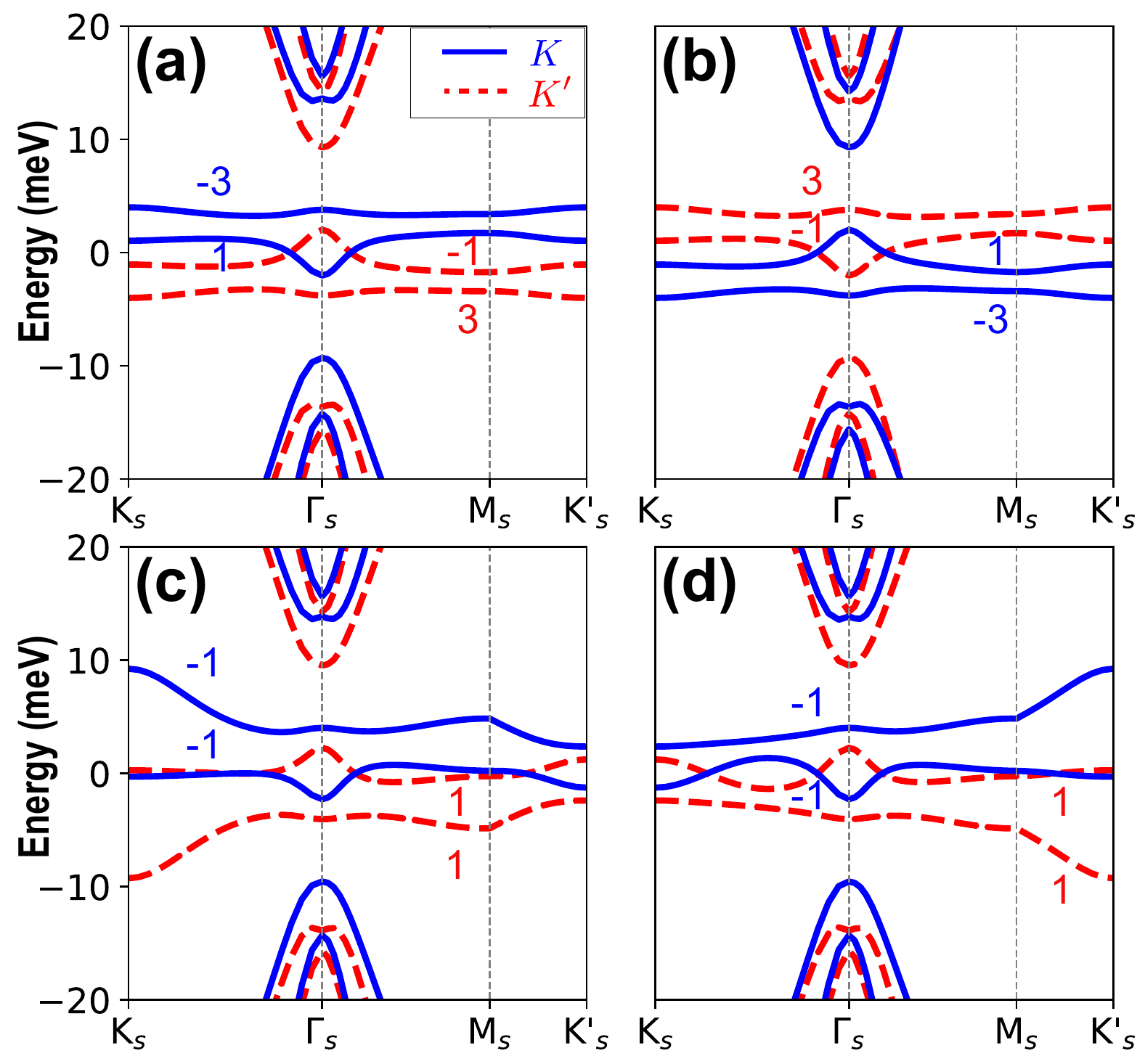}
\caption{\label{fig:fig2} Floquet band structures  of the AB-BA TDBG at $\theta=1.05^\circ$. (a) Left CPL with $\Delta_\Omega=4\,\mathrm{meV}$, $\Delta_E=0\,\mathrm{meV}$. (b) Right CPL with $\Delta_\Omega=-4\,\mathrm{meV}$, $\Delta_E=0\,\mathrm{meV}$. (c)  Left CPL with $\Delta_\Omega=4\,\mathrm{meV}$, $\Delta_E=3.5\,\mathrm{meV}$. (d) Left CPL with $\Delta_\Omega=4\,\mathrm{meV}$, $\Delta_E=-3.5\,\mathrm{meV}$.}
\end{figure}

In high frequency limit, according to the Floquet band theory, a reasonable approximation is to expand the time dependent Hamiltonian in Eq.~\eqref{TimeH}  to the first order of $1/\Omega$ \cite{Eckardt2015, Mikami2016},
\begin{equation}\label{Fhamiltonian}
H_{K, M+N}^{\mathrm{F},\eta,\eta'} = H_{K, M+N}^{(0),\eta,\eta'} + \Delta_{\Omega}I_{M+N}\otimes \sigma_z + U,
\end{equation}
where $H_{K, M+N}^{(0),\eta,\eta'}$ is the TMG Hamiltonian without irradiation. An important message is that the CPL give rise to an additional mass term in Eq.~\eqref{Fhamiltonian}. Here, $I_{M+N}$ is an identity matrix of order $M+N$ and $\Delta_\Omega =\xi (e v_FA_0)^2/\hbar\Omega$.  The Hamiltonian equation (\ref{TimeH}) used in this study is the continuum model, which is meant to capture the low energy physics of TMG system at small twisted angles~\cite{Liu2019}. Detailed features such as domain formation and inhomogeneity of twisting angles brought by lattice relaxation have not been included. Instead, we adopt $w_{AA} < w_{AB}$ to partially account for the lattice relaxation effect~\cite{Chebrolu2019, Carr2019a, Tarnopolsky2019}. For CPL, we only consider its effect on the modification of the electronic band structure. In order to get effective Hamiltonian equation (\ref{Fhamiltonian}), we assume the applied CPL frequency is in the off-resonant limit so that high frequency expansion can be used\cite{Oka2009, Mikami2016}.

The parameters for the MG used in this paper is adapted from Ref.~\cite{Chebrolu2019} and we record them here for convenience: $t_0\equiv 2\hbar v_F/\sqrt{3}a_0 = -3.1$\,eV is the intra-layer nearest neighbour hopping  with $a_0=2.46$\,\AA \, and $v_F \approx 1.0\times10^6\,\mathrm{m/s}$; $w_{\mathrm{AB}}=0.12\,\mathrm{eV}$, $w_{\mathrm{AA}}=0.098\,\mathrm{eV}$ and $t_{\bot}\approx 3w_{AB}=0.36\,\mathrm{eV}$. For a $4$\,meV $\Delta_{\Omega}$ with  photon energy $\hbar\Omega=1.5\,\mathrm{eV}$, the electrical field strength of the CPL is about $1.76\times 10^3 \,\mathrm{KV/cm}$ and the intensity is about $4.11\times 10^{9} \,\mathrm{watt/cm^2}$, which should be feasible in experiments  with  ultrafast laser technique \cite{Mourou1998,Sato2019, Mciver2019,Vogl2020a}. Since the central flat bands have a rather small band width, the light frequency needed to reach the off-resonant limit is significantly reduced. A choice of 1.5eV light frequency should let us in the off-resonant regime~\cite{Katz2020, Topp2019}.

\begin{figure}[tbp]
\includegraphics[scale=0.52]{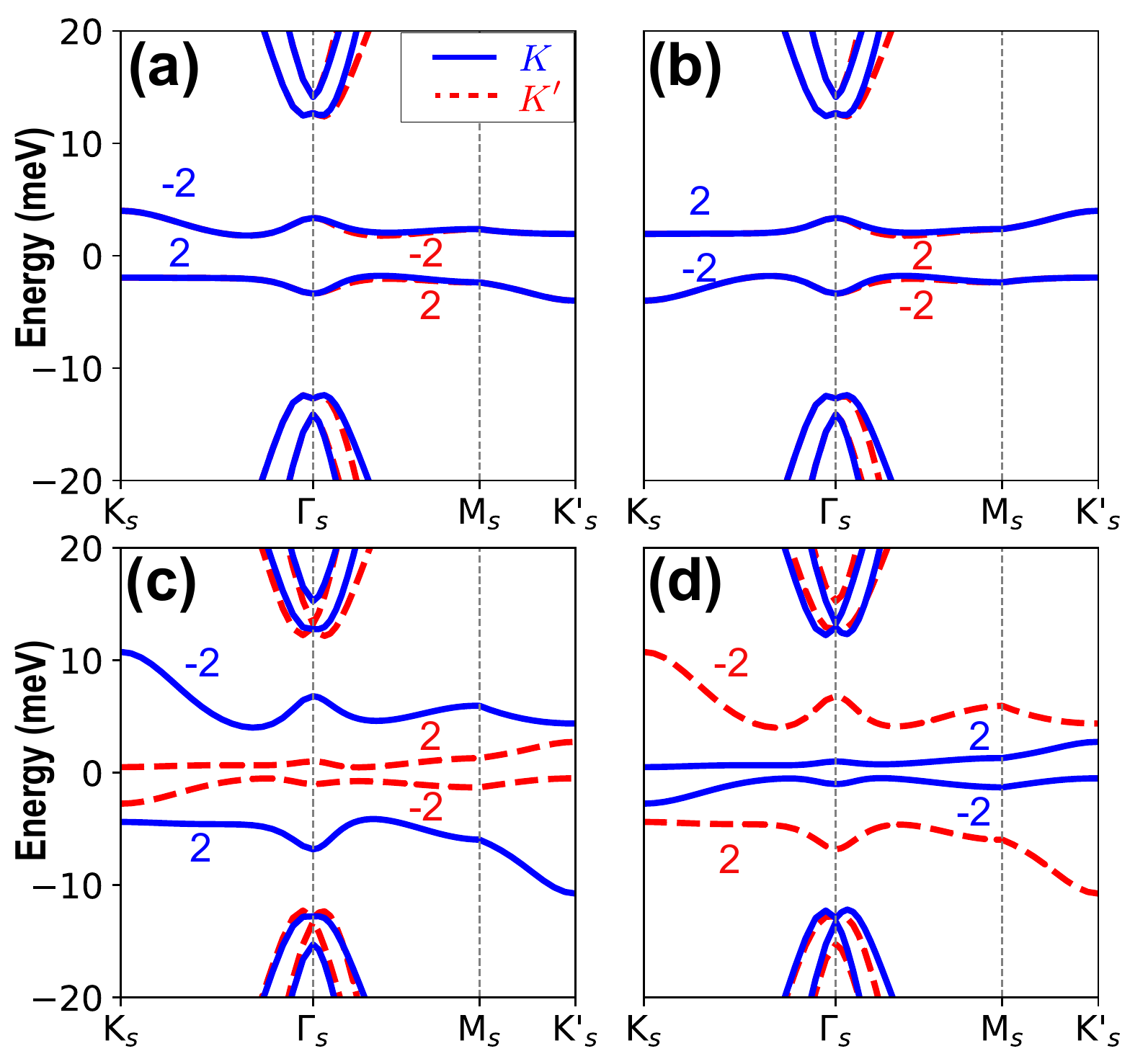}
\caption{\label{fig:fig3}  Floquet band structures  of the AB-AB TDBG at $\theta=1.05^\circ$. (a) Left CPL with $\Delta_\Omega=4\,\mathrm{meV}$, $\Delta_E=0\,\mathrm{meV}$. (b)Right CPL with $\Delta_\Omega=-4\,\mathrm{meV}$, $\Delta_E=0\,\mathrm{meV}$. (c) Left CPL with $\Delta_\Omega=4\,\mathrm{meV}$, $\Delta_E=4.5\,\mathrm{meV}$. (d). Left CPL with $\Delta_\Omega=4\,\mathrm{meV}$, $\Delta_E=-4.5\,\mathrm{meV}$.}
\end{figure}

\section{Floquet Chern Flat bands in TDBG}
We take the TDBG  as a paradigm to illustrate the CPL effects on TMG. We first discuss the case of AB-BA TDBG, as an example of o-TMG. The most remarkable result is that it has two separated and valley-split Floquet Chern flat bands with nonzero total Chern number.  Figs.~\ref{fig:fig2} (a) and \ref{fig:fig2} (b) show the Floquet band structures of AB-BA TDBG  at $\theta=1.05^\circ$ around the first magic angle under left and right CPL, respectively. In each moir\'{e} valley, we have two nearly flat bands near the charge neutrality point, which are separated by a gap about $1.5$ meV. Note that the gap is due to the applied CPL and can be increased with larger $\Delta_\Omega$, while in pristine TDBG the two flat bands are  touched at the $K_s$ and $K_s^{'}$ points [marked in Fig.~\ref{fig:fig1}(d)].
The energy of the two Floquet flat bands are valley-dependent.  As shown in Fig.~\ref{fig:fig2}(a), with left CPL, the valence band in the moir\'{e} $K'$ valley (red dashed lines) is lower than that in the $K$ valley (blue solid lines).  The case becomes opposite when a right CPL is applied, see Fig.~\ref{fig:fig2} (b). It means that, in AB-BA TDBG, a valley-polarized flat band can be selectively generated by CPL with different polarization.  Meanwhile, these flat bands are topologically nontrivial, for which the valley Chern number and the total Chern number are both nonzero. For example, in Fig.~\ref{fig:fig2}(a), the valley Chern numbers of the valence band in $K$ and $K'$ valleys are $C_{Lv}^{K}=1$ and $C_{Lv}^{K'}=3$, respectively. Thus, the total Chern number of the valence band is $C^{tot}_{Lv}=C_{Lv}^{K}+C_{Lv}^{K'}=4$. The nonzero total Chern number is allowed here since time reversal symmetry is broken by CPL. If the polarization of CPL is changed, the total Chern number changes sign, see Fig.~\ref{fig:fig2}(b). The topological properties of the flat bands can be further adjusted by a gate-induced perpendicular electric field represented by $\Delta_E$. In Figs.~\ref{fig:fig2}(c) and \ref{fig:fig2}(d), with $\Delta_E=\pm3.5$ meV,  the valley Chern numbers in both valleys change.

As shown in Fig.~\ref{fig:fig1} (f), the moir\'{e} $K$ ($K'$) valley of o-TMG is formed by mixing two opposite chirality MG valleys $K^+_M$ and $K^-_N$ ($K'^-_M$ and $K'^+_N$). Specifically, the states at $K_s$ point are mainly formed  by electrons from the positive chirality MG valleys, while that at $K'_s$  point are from negative chirality MG valleys.
We know that, for irradiated MG, the  CPL induced  mass terms have opposite sign in the two valleys depending on their chirality, while the gate voltage induced ones have the same sign \cite{Qu2017, Li2018}. Hence, CPL induced mass terms have opposite signs at $K_s$ and $K'_s$, while that induced by electric field have the same sign. Consequently, due to the competition between $\Delta_\Omega$ and $\Delta_E$, when gradually increasing the gate voltage, $\Delta_E>0$ for example, the gap at $K_s$  point enlarges, while the gap at $K'_s$  point first diminishes and then reopens, resulting in the valley Chern numbers change in both moir\'{e} valley  and the imbalanced gaps open at $K_s$ and $K'_s$ points.


Now, we turn to the case of AB-AB TDBG, \textit{i.e.}~an example of s-TMG. First, with only CPL, we cannot split the valley degeneracy. The Floquet bands of AB-AB TDBG under left and right CPL are shown in Figs.~\ref{fig:fig3}(a) and \ref{fig:fig3}(b), respectively. With left (right) CPL, we get two separated Floquet flat bands with $C_{Lc}^{K}=-2$, $C_{Lv}^{K}=2$ ($C_{Rc}^{K}=2$, $C_{Rv}^{K}=-2$), \textit{etc}. There are two obvious differences between the AB-AB and AB-BA cases.  One is that the energy of flat bands in two moir\'{e} valleys are mixed in the AB-AB situation, and the other is that their corresponding valley Chern numbers are different.  However, the total Chern number of the conduction or valance flat bands are the same, \textit{e.g.}~$C^{tot}_{Lv}=4$ as in Fig.~\ref{fig:fig3}(a) and Fig. ~\ref{fig:fig2}(a).

Interestingly, the valley degeneracy in AB-AB TDBG can be further lifted by an additional vertical electric field.   In Fig.~\ref{fig:fig3}(c), we apply a left CPL ($\Delta_\Omega=4$ meV) and a perpendicular electric field ($\Delta_E=4.5$ meV). The applied electric field induces a valley-dependent energy shift, and now the valence band in $K$ valley is lower than that in $K'$ valley. In contrast, when we reverse the direction of electric field, the valence band in the  $K'$ valley has lower energy, see in Fig.~\ref{fig:fig3}(d).  So, in the AB-AB TDBG, we can also selectively get a valley-polarized Floquet Chern flat band via the combination effect of CPL and vertical electric field. The electric field can also change the valley Chern number, see in Figs.~\ref{fig:fig3}(c) and \ref{fig:fig3}(d). For example, with large enough positive $\Delta_E$ [Fig.~\ref{fig:fig3}(c)], the valley Chern numbers in $K$ valley is invariant, while that of the $K'$ valley is changed. Consequently, the total Chern number of each flat band becomes zero, which indicates a transition from quantum anomalous hall state to quantum valley hall state.

As shown in Fig.~\ref{fig:fig1}(e), for the s-TMG, the $K$  mori\'e valley is formed by the $K_M^+$ and $K_N^+$ MG valleys, while $K'$ moir\'{e} valley is formed by $K_N^{'-}$ and $K_M^{'-}$. So, with similar reason as before, the mass term induced by CPL now will  be opposite in sign for the two mori\'e valleys, and the electrical field induced  one   have the same sign at the two mori\'e valleys. When gradually increasing the  gate voltage, the gap at one mori\'e valley further increases and the gap at the other first diminishes and then reopens, resulting in the moir\'{e}-valley-contrasting gap and the valley Chern number change only in  one mori\'e  valley. We see that the valley polarization mechanisms for o-TMG and s-TMG are different, and their different responses to CPL are caused by the interplay between the valley selective nature of CPL and the stacking chirality induced distinct MG valleys combination.

A valley-polarized flat band has many novel features, such as orbital ferromagnetism, magneto-optical effect, \textit{etc} \cite{liujianpeng2020}. Now, we have demonstrated that the valley polarization in TMG can be selectively generated by Floquet engineering with CPL, so that many of these novel properties may be controlled by CPL irradiation. For example, we predict that CPL is able to generate and switch orbital ferromagnetism in TMG. Meanwhile, we anticipate that if superconductivity is formed, it may favor an exotic FFLO states \cite{Fulde1964, Larkin1965}.

\begin{table}[h]
	\caption{\label{tab:table1}%
		Valley Chern numbers of the (M+N)-layer TMG. The upper part is for s-TMG and lower part is for o-TMG.
	}
	\begin{ruledtabular}
		\begin{tabular}{cccccccccc}
			\textrm{M}&
			\textrm{N}&
			\textrm{$C_{Lc}^K$}&
			\textrm{$C_{Lv}^K$}&
			\textrm{$C_{Lc}^{K'}$}&
			\textrm{$C_{Lv}^{K'}$}&
			\textrm{$C_{Rc}^{K}$}&
			\textrm{$C_{Rv}^{K}$}&
			\textrm{$C_{Rc}^{K'}$}&
			\textrm{$C_{Rv}^{K'}$}\\
			\colrule
			2 & 2 & -2 & 2 & -2 & 2& 2 & -2 & 2& -2\\
			2 & 3 & -2 & 3 & -3 & 2& 3 & -2 & 2 & -3\\
			3 & 3 & -3 & 3 & -3 & 3& 3 & -3 & 3 & -3\\
			2 & 4 & -2 & 4 & -4 & 2& 4 & -2 & 2 & -4\\
			M & N & -M & N & -N & M& N & -M & M  &-N \\
			\colrule
			2 & 2 & -3 & 1 & -1 & 3 & 1 & -3 & 3& -1\\
			2 & 3 & -4 & 1 & -1 & 4 & 1 & -4 & 4& -1 \\
			3 & 3 & -5 & 1 & -1 & 5 & 1 & -5 & 5 & -1\\
			2 & 4 & -5 & 1 & -1 & 5 & 1 & -5 & 5 & -1\\
			M & N & -M-N+1 & 1 & -1 & M+N-1& 1 & -M-N+1 & M+N-1 & -1
		\end{tabular}
	\end{ruledtabular}
\end{table}

\section{Floquet Chern flat bands of (M+N)-layer TMG}
The main features of the Floquet flat bands in a general (M+N)-layer TMG are quite like that in TDBG, \textit{i.e.}~two separated Chern flat bands near the fermi level. It is because that all the  ABC-stacked MGs have similar band structure. Most importantly, the CPL induced valley splitting is valid for all the TMGs. The main difference is the topological features of the Floquet flat bands. Here, we find a valley Chern number hierarchy rule of the Floquet flat bands in the (M+N)-layer TMG:
\begin{equation}\label{CS}
C_{\xi \zeta}^{\tau} = -\frac{1}{2} \left\{\xi\zeta(M+N)+\tau\left[(M-1)-s(N-1)\right]\right\}
\end{equation}
where $s=1$ is for s-TMG and $s=-1$ for o-TMG; $\zeta$ is the band index for the conduction ($\zeta=+1$) and valence ($\zeta=-1$) band; $\tau$ denotes the moir\'{e} valley and $\xi$ represents the polarization of CPL as mentioned before.

Equation \eqref{CS} is summarized from our numerical results (see the Table \ref{tab:table1}). For a theoretical proof, see Appendix B. Interestingly, the total Chern number of either conduction or valence band is $-\xi\zeta (M+N)$,  which is  irrelevant to the stacking chirality. The non-zero total Chern number is not so surprise here because time reversal symmetry is broken by CPL. Another rule of Chern number is about the total Chern number of the conduction and valence bands in each moir\'{e} valley. In  s-TMG, the sum of valley Chern number in a single moir\'{e} valley equals $-\tau(M-N)$, while for the o-TMG, it is $-\tau(M+N-2)$. This rule is irrelevant with polarization of light, and originates from the fact that, within the parameter regime used in this article, the two central flat bands are separated from other higher energy bands. Therefore, the sum of the two central flat bands within one mori\'e valley is the same as in the case without the CPL driven ~\cite{Liu2019}.

\section{Summary and Outlook}
In summary, we have studied the Floquet flat bands in the TMG systems. We illustrate that CPL can selectively produce a valley polarized Floquet Chern flat band with tunable large Chern number. It offers a feasible way to manipulate the valley degree of freedom for the moir\'{e} flat bands, and provides a new platform to study  moir\'{e} valleytronics~\cite{Hu2018}.  We predict that many of the valley-related phenomena in TMG can be switched by CPL, \textit{e.g.}~optical controlled orbital ferromagnetism and valley Hall effect~\cite{Sharpe2019,Serlin2019, Tschirhart2020, Komatsu2018}. We reveal that CPL effects here strongly rely on the  stacking chirality of TMG, and we also find a Chern number hierarchy rule for the Floquet flat bands in TMG.  Here we note that the Chern number hierarchy rule listed in Table I is derived using the simple effective Hamiltonian equation (\ref{Fhamiltonian}), assuming the twisted angle is near the first magic angle and light intensity is low (small $\Delta_\Omega$). In real TMG materials, a complete inclusion of lattice relaxation may have some important impacts on this rule\cite{Haddadi2020}. Experimentally, TDBG has been made in laboratory and recently light induced anomalous Hall effect in monolayer graphene has finally been observed~\cite{cao2020,liu2020tunable,cheng2020,Burg2019,Mciver2019}. We think the different responses to CPL for TDBG with different stacking chiralities can be readily tested within current experimental techniques.

The study in this article is in the off-resonant regime, where the high frequency expansion approximation is used. The precision of this description depends on the details of state preparation. For example, to resemble a topological insulator in equilibrium, electrons should mostly populate within a chosen Floquet replica. This can be achieved by connecting to a bath with particular engineered density of states, which is used to suppress the photon-assistant tunneling process to other Floquet replicas\cite{Iadecola2015, Seetharam2015}. For a closed system without coupling to any reservoirs, to prepare the topological nontrivial state from the trivial one, an optimal choice of ramp speed for the driving amplitude is needed. Generally speaking, the ramp speed should be neither too fast nor too slow. If too fast, the resulting state is not predicted well by the static effective Hamiltonian $H_{\text{eff}}$ \cite{DAlessio2015}; while if it is too slow, heating effect brought by interactions may completely destroy the interesting topological features\cite{Ho2016, Weinberg2017}. It is estimated that the optimal ramp speed to be $t_{\text{heat}}^{-z/(z+d)}$, where $t_{\text{heat}}$ is the heating timescale, $d$ is the dimension and $z$ is the critical exponent of the correlation length \cite{Ho2016}. Experimental measurements should also be done before the deleterious effects by heat set in. The heating time rapidly increases with increasing driving frequency\cite{Abanin2015, Kandelaki2018}. For the off resonant case considered here, there should be an extended time window for the interesting topological features to be measured. The detailed study of state preparation for driving TMGs is beyond the scope of this paper and we call for future studies.

\emph{Note added.} We  note an independent theoretical work discussing the Floquet engineering in twisted double bilayer graphene \cite{RodriguezVega2020}.
\begin{acknowledgments}
Ming Lu thanks Hua Jiang, Robert Joynt and Jianpeng Liu for helpful discussions. This work is supported  by the National Natural Science Foundation of China (Grants No.~11534001, 11874160, 11274129, 11874026, 61405067), and the Fundamental Research Funds for the Central Universities (HUST: 2017KFYXJJ027), and NBRPC (Grants No. 2015CB921102).
\end{acknowledgments}

\appendix

\section{EFFECTIVE HAMILTONIAN UP TO SECOND ORDER}
We derive the effective Floquet Hamiltonian up to second order in this appendix. According the Brillouin-Wigner expansion method \cite{Mikami2016}, the zeroth to second order effective Hamiltonian writes:
\begin{align}
	H^{(0)} &= H_0 
	\\
	H^{(1)} &= \sum_{\{n_i\neq 0\}} \frac{H_{-n_1} H_{n_1}}{n_1\hbar\Omega}
	\\
	H^{(2)} &= \sum_{\{n_i\neq 0\}} \frac{H_{-n_1}H_{n_1-n_2}H_{n_2}}{n_1n_2(\hbar\Omega)^2}  - \frac{H_{-n_1}H_{n_1}H_{0}}{n_1^2(\hbar\Omega)^2}
\end{align}
where $H_n=\frac{1}{T}\int_0^{T} H(t) e^{\mathrm{i}n\Omega t}\mathrm{d} t$. For the continum model described by Equation (\ref{TimeH}), only $H_{-1}$, $H_{0}$ and $H_{1}$ survives. In the following, we assume left circularly polarized light and mori\'{e} K valley for concreteness:
\begin{align}
	H_0 &= H_{K, M+N}^{(0), \eta, \eta'}\\
	H_{-1} &= ev_F A_0 \mathrm{diag}\left[ \sigma_{+}, \sigma_{+}, \cdots, \sigma_{+} \right]\\
	H_{1} &= ev_F A_0 \mathrm{diag}\left[ \sigma_{-}, \sigma_{-}, \cdots, \sigma_{-} \right]
\end{align}
where $H_{K, M+N}^{(0), \eta, \eta'}$ is the TMG Hamiltonian without irradiation. The first order correction is:
\begin{align}
	H^{(1)}& = \frac{1}{\hbar \Omega} \left[H_{-1}, H_{1}\right]\\ 
	&=\frac{(ev_F A_0)^2}{\hbar\Omega}\mathrm{diag}\left[ \sigma_{z}, \sigma_{z}, \cdots, \sigma_{z} \right]
\end{align}
The second order correction is:
\begin{align}
H^{(2)} = \frac{1}{(\hbar\Omega)^2} \left[(H_{-1}H_0 H_1+H_1 H_0 H_{-1})\right] \nonumber\\ -\frac{1}{(\hbar\Omega)^2}(H_{-1}H_1+H_1H_{-1})H_0
\end{align}
It is easy to see $H_{-1}H_1+H_1 H_{-1}$ is proportional to the identity matrix. Further, note that $\sigma_{-}\sigma_x\sigma_{+}$, $\sigma_{+}\sigma_y\sigma_{-}$, $\sigma_{+}^2$, $\sigma_{-}^2$ all equal to zero. Therefore, the sum of first two terms has only two nonzero entries which are proportional to $w_{AA}$. Explicitely:
\begin{align}
	H^{(2)} = \frac{(ev_FA_0)^2}{(\hbar\Omega)^2} \left[\begin{pmatrix}
	O& \mathbb{W}\\
	\mathbb{W}^\dagger & O
	\end{pmatrix} - H_0\right]
\end{align} 
where $\mathbb{W} = E_{M\times N}\otimes W(r)$, $E_{M\times N}$ is a $M\times N$ matrix with only one nonzero entry $E_{M\times N}(M, 1)=1$ and $W(r)= w_{AA}\sigma_0\sum_{n=1}^3 e^{-i\bm{q}_n\cdot \bm{r}}$. For our choice of light frequency and amplitude in Fig.~\ref{fig:fig2}, we have $\Delta_\Omega = \frac{(ev_FA_0)^2}{\hbar\Omega} = 4 \text{meV}$ and $\hbar \Omega = 1.5\text{eV}$, therefore $\alpha \equiv \frac{(ev_FA_0)^2}{(\hbar\Omega)^2}=\frac{\Delta_\Omega}{\hbar \Omega}\approx 2.7\times 10^{-3} \ll 1$. Including second order correction amount to change $w_{AA}$ to $\frac{w_{AA}}{1-\alpha}$ in $H_0$ and multiply $(1-\alpha)$ in front of $H_0$. This correction is very small and won't affect the physics we are discussing. Therefore, in the main text we consider only the first order correction.

\section{PROOF THE CHERN NUMBER HIERARCHY RULE}
In this appendix, we prove the Chern number hierarchy rule in equation (\ref{CS}). As described in the main text, the sum of conduction and valence band Chern numbers within a single valley satisfies:
\begin{equation}\label{A1}
	C_{\xi \zeta}^{\tau}+C_{\xi -\zeta}^{\tau} = -\tau[(M-1)-s(N-1)]
\end{equation}
where $s=\pm1$ is for s-TMG (o-TMG); $\zeta=\pm 1$ is the conduction (valence) band index; $\tau=\pm 1$ denotes the moir\'{e} K (K') valley and $\xi=\pm 1$ represents the left (right) CPL. This relation is proved in the case where no driving is present and the central flat bands are degenerate at $K_s$ and $K'_s$ points\cite{Liu2019}. Since we are considering the case where driving induced $\Delta_\Omega$ is small and the central flat bands do not touch with the higher energy bands, this relation is unchanged.

For the sum of conduction or the valence bands among the two mori\'{e} valleys, we have:
\begin{equation}
	C_{\xi \zeta}^{\tau}+C_{\xi \zeta}^{-\tau} = -\xi\zeta(M+N)
\end{equation}
This relation can be physically understood as follows: with the CPL dirven: the top M-layer lowest Floquet energy band has a Chern number $-\xi\zeta M$ and the bottom N-layer lowest Floquet energy band has a Chern number $-\xi\zeta N$ \cite{Li2018}; since the mori\'{e} conduction (valence) flat bands is formed by coupling these lowest conduction (valence) M-layer and N-layer Floquet bands near Dirac point (as schematically shown in Fig. \ref{fig:fig1} (e) and \ref{fig:fig1}(f)), the sum of their Chern numbers among two mori\'{e} valleys is therefore the sum of the Chern numbers of their constitutions.

The third relation is between the condution flat band in one mori\'{e} valley and valence flat band in the other mori\'{e} valley, which reads:
\begin{equation} \label{A3}
	C_{\xi \zeta}^{\tau}=-C_{\xi -\zeta}^{-\tau}
\end{equation}
To prove this relation, we first define a chiral operator $S=I_{M+N}\otimes \sigma_z$, the effective Hamiltonian for the two valleys write as $H_{\tau, M+N} = H_{\tau,M+N}^{(0)} + \tau S \Delta_\Omega$, with $H_{\tau, M+N}^{(0)}$ represent the undriven Hamiltonian in mori\'{e} $\tau$ valley, see equation (\ref{Fhamiltonian}) and its discussions in the main text. In the chiral limit where $w_{AA}=0$, $S^{-1}H_{\tau, M+N}^{(0)} S = -H_{\tau, M+N}^{(0)}$, therefore:
\begin{equation}
S^{-1}H_{+, M+N} S = -H_{-, M+N}^{*}
\end{equation}
and $H_{-,M+N}(\vec{k})= -S^{-1} H_{+,M+N}^{*}(-\vec{k}) S$. The energy bands and the wave functions of the two valleys has the relation:
\begin{equation}
E_{-n\vec{k}}^{(+)}=-E_{n-\vec{k}}^{(-)},\qquad \psi_{-n,\vec{k}}^{(+)}(\vec{r})=S\psi_{n-\vec{k}}^{(-)*}(\vec{r})
\end{equation}
where $E_{n\vec{k}}^{(\tau)}$ and $\psi_{n,\vec{k}}^{(\tau)}(\vec{r})$ are the n-th energy bands and wave function in mori\'{e} $\tau$ valley, respectively. With this, we have:
\begin{equation}
\vec{A}_{n\vec{k}}^{(-)} = i\left<\psi_{n\vec{k}}^{(-)}(\vec{r})\mid\frac{\partial}{\partial \vec{k}}\mid \psi_{n\vec{k}}^{(-)}(\vec{r})\right>
 = \vec{A}_{-n-\vec{k}}^{(+)}
\end{equation}
The Berry curvature is derived from Berry connection:
\begin{equation}
\Omega_{n\vec{k}}^{(-)}=\frac{\partial}{\partial k_x} A_{n\vec{k},y}^{(-)} - \frac{\partial}{\partial k_y} A_{n\vec{k},x}^{(-)}=-\Omega_{-n-\vec{k}}^{(+)}
\end{equation}
where $\vec{A}_{n\vec{k}}^{(\tau)}$ and $\Omega_{n\vec{k}}^{(\tau)}$ are the Berry connection and Berry curvature of n-th band in the mori\'{e} $\tau$ valley. Take $n=\pm 1$ which represent central flat bands and integrate over the mBZ, we get equation (\ref{A3}). 

The flat bands around the magic angle is almost perfectly flat in the chiral limit \cite{Tarnopolsky2019, Liu2019}, as we gradually turn on $w_{AA}$ to the final value and breaks chiral symmetry, the bands become dispersive. Since the central flat bands do not touch with other bands during this process, the valley Chern numbers are unchanged and this relation still holds. 

With relations (\ref{A1})-(\ref{A3}), we get the equation (\ref{CS}) in the main text.


\bibliography{reference}

\end{document}